\documentclass[12pt, 
]{iopart}
\usepackage{xspace}
\usepackage[nolist,nohyperlinks]{acronym}
\usepackage{graphicx}
\usepackage{titlesec}
\usepackage{textcomp}
\usepackage[obeyDraft]{todonotes}
\usepackage[separate-uncertainty = true,
			multi-part-units = repeat
			]{siunitx}
\usepackage{setspace}
\usepackage{textcomp}
\usepackage[left,switch]{lineno}
\usepackage[sort&compress,numbers,square]{natbib}
\usepackage{pdfpages}
\titleformat*{\subsection}{\small \bfseries}

\begin{acronym}
   \acro{BCA}{binary collision approximation}
   \acro{bftem}[BF-TEM]{bright-field \acl{TEM}}
   \acro{cmos}[CMOS]{complementary metal oxide semiconductor}
   \acro{dpa}[dpa]{displacements per atom}
   \acro{EBDW}{electron beam direct write}
   \acro{eftem}[EFTEM]{energy-filtered \acl{TEM}}
   \acro{fft}[FFT]{Fast Fourier transform}
   \acro{FIB}{focused ion beam}
   \acro{GAA}{gate-all-around}
   \acro{HIM}{helium ion microscope}
   \acro{HT}{high temperature}
   \acro{MC}{Monte-Carlo}
   \acro{RT}{room temperature}
   \acro{SET}{single electron transistor}
   \acro{TEM}{transmission electron microscopy}
   \acro{VLSI}[VLSI]{very-large-scale integration}
\end{acronym}

\newcommand{\neion}{Ne$^+$}

\begin{document}
	
\onehalfspacing{}

\title[]{Morphology modification of Si nanopillars under ion irradiation at elevated 
		temperatures: plastic deformation and controlled thinning to 10\,nm}
\author[1]{Xiaomo Xu, Karl-Heinz Heinig, Wolfhard M\"oller, Hans-J\"urgen Engelmann,  
        Nico Klingner, Johannes von Borany 
        and Gregor Hlawacek\thanks{}}
\address{Institute of Ion Beam Physics and Materials Research, Helmholtz-Zentrum 
		Dresden-Rossendorf, Bautzner Landstrasse 400, 01328 Dresden, Germany}

\author[2]{Ahmed Gharbi, Raluca Tiron}
\address{CEA-Leti, Rue des Martyrs 17, 38054 Grenoble, France}
\ead{g.hlawacek@hzdr.de}

\vspace*{1\baselineskip}
\textbf{\today}
\vspace*{1\baselineskip}

\listoftodos{}

\begin{abstract}
   Si nanopillars of less than 50\,nm diameter have been irradiated in a helium ion microscope 
   with a focused Ne$^+$ beam. The morphological changes due to ion beam irradiation at room 
   temperature and elevated temperatures have been studied with the transmission electron microscope. 
   We found that the shape changes of the nanopillars depend on irradiation-induced amorphization 
   and thermally driven dynamic annealing. While at room temperature, the nanopillars evolve to 
   a conical shape due to ion-induced plastic deformation and viscous flow of amorphized Si, 
   simultaneous dynamic annealing during the irradiation at elevated temperatures prevents 
   amorphization which is necessary for the viscous flow. Above the critical temperature of 
   ion-induced amorphization, a steady decrease of the diameter was observed as a result of the dominating 
   forward sputtering process through the nanopillar sidewalls. Under these conditions the nanopillars 
   can be thinned down to a diameter of \SI{\sim{}10}{\nano\meter} in a well-controlled manner. 
   A deeper understanding of the pillar thinning process has been achieved by a comparison of 
   experimental results with 3D computer simulations based on the binary collision approximation.
\end{abstract}

\vspace{2pc}

\submitto{\SST}

%
\maketitle
%
%

\section{Introduction}

While the development of new device architectures and computing strategies should be disruptive 
to enable a rapid progress in the field, the underlying fabrication technology ideally builds upon 
existing knowledge to enable a quick and seamless transition to the new data processing regime.
Examples for this approach towards the development of future computing units are new low power devices such as \acp{SET} or Si-based quantum computing 
strategies~\cite{Lee2014, Morse2017}.
The existing \ac{cmos} technology can handle today's demands on feature size and shape, but new 
architectures require new---ideally \ac{cmos}-compatible---fabrication processes to meet the upcoming 
challenges. Here, we present a \ac{cmos}-compatible ion beam based fabrication process for 
sub \SI{15}{\nano\meter} pillars with the potential for \ac{GAA} \ac{SET} device architectures~\cite{Xu2018} 
and other similar technologies that require three dimensional building blocks with lateral length 
scales in the few-\si{\nano\meter} regime. We investigate the achievable diameter reduction for 
sub-\SI{50}{\nano\meter} pillars with a height of \SI{70}{\nano\meter} and also propose a model 
for the ion beam based mechanism.

The interactions between ion beams and Si-based material systems have been extensively studied in 
the last decades due to the ubiquitous and manifold applications of ion beams for device fabrication. 
Examples include the doping of transistor active regions~\cite{Mayer1973, Campisano1993} mostly 
using broad ion beams and milling of nanostructures~\cite{Schuhrke1992, Pekin2016} using \acp{FIB}. 
Side effects of ion beam processing of Si include defect accumulation and amorphization. 
Studies---mostly with unstructured bulk materials---to understand the mechanism and to avoid these 
effects have been carried out with various ion species, energy ranges, target structures 
etc.~\cite{Zhong2004} and are comprehensively summarized in review papers~\cite{Pelaz2004, Wesch2012} 
and textbooks~\cite{Wesch2016,Nastasi2006}.

Ion irradiation at elevated temperatures has been considered as a promising technique to mitigate ion 
beam induced damages and amorphization of the substrate during the irradiation processes~\cite{Williams1992,Goldberg1995,Stanford2016}.
For most semiconductors, a finite temperature T$_\mathrm{c}$---lower than the temperature for 
epitaxial recrystallization~\cite{Crowder1970,Mayer1968}\footnote{for Si this is 
in the range of \SIrange{550}{600}{\celsius}}---exists at which amorphization is prevented during 
irradiation. This is related to out-diffusion of vacancies from the ion track region, and is described in the so called out-diffusion theory~\cite{Morehead1970, Dennis1978}.
According to this theory, at any finite temperature, a thermally driven dynamic annealing process, 
characterized by Si interstitial-vacancy recombination, 
competes with the amorphization caused by the incident ion.
When the substrate temperature T $\ll$ T$_\mathrm{c}$, the ion damage prevails and Si 
undergoes continuous amorphization. Once the temperature exceeds T$_\mathrm{c}$ the dynamic annealing will 
recover the amorphized pocket of each incident ion, thus preventing the amorphization.
The process is governed by ion mass, target temperature and---to a lesser extent---flux and ion energy.
Here, we utilize this dynamic annealing process at slightly elevated temperatures to prevent the 
amorphization of nanostructures during ion beam irradiation. 

In the case of the nanostructures used in this work the length scale of the ion collision cascade 
becomes comparable to the structure size which results in unanticipated effects. 
Such effects can arise from changes in the distribution of the deposited energy.
These changes are a result of the truncation of the collision cascade by the nanostructure. 
This reduces the amount of energy deposited into the nanostructure and changes the distribution of 
the deposited energy as the final low particle energy part of the cascade is missing. 
However, this part is characterized by a high relative nuclear energy loss as compared to the part 
of the cascade closer to the impact point. 
In addition, new processes such as backside or forward sputtering can change the stability of the 
nanostructure during ion beam irradiation~\cite{Nietiadi2014}. 

In this work we employ \ac{FIB} and broad-beam irradiation in the few tens of \si{\kilo\electronvolt} 
range to shrink the diameter of few-tens of \si{\nano\meter} pillars down to nearly \SI{10}{\nano\meter}.
Specifically we use \neion{} irradiation at \SI{25}{\kilo\electronvolt} under normal incident in a 
\ac{HIM}~\cite{Hlawacek2013c,Hlawacek2016}.
In the \ac{HIM} the sample is heated in-situ using a home-built heater stage that can be loaded through the load lock of 
the Carl Zeiss Orion NanoFAB\@. Although the \ac{HIM} has a lateral resolution of less than 
\SI{2}{\nano\meter} when using \neion{}, we 
scan the beam over a set of nanostructures to emulate a broad beam irradiation, to which we also 
compare the results at the end of the manuscript. The benefit of this approach is that a few 
pillars from the same sample chip can be irradiated at different fluences and/or at different temperatures.
Possible morphological or structural changes can than directly be compared in the subsequent HIM 
imaging step at RT\@. This characterization has been performed using high resolution \ac{HIM} and \ac{TEM} to 
obtain information on the morphology and crystallinity of the obtained nanopillars, respectively.

The recently developed Monte Carlo simulation program TRI3DYN~\cite{Moeller2014} is used to perform a 
fully dynamic 3D simulation of the irradiation process. 
This is complemented by sputter yields and distributions of sputtered particles extracted from the 
static collision simulation program TRI3DST~\cite{Nietiadi2014,Moeller2016}. 
The simulated results are compared with experimental findings and help with the understanding of the 
underlying processes.

To demonstrate the possibility for upscaling of this method we also employ Si$^+$ broad beam 
irradiation. Here, always the entire sample is irradiated and multiple samples have to be used 
to investigate the influence of temperature and ion fluence. Both local and broad beam based 
irradiation suggest a versatile and \ac{cmos}-compatible fabrication method of sub-\SI{10}{\nano\meter} 
diameter vertical Si nanostructures at slightly elevated but still \ac{cmos}-compatible temperatures.

\section{Methods}

Silicon nanopillars have been fabricated via patterning of Si-rich anti-reflective coating (SiARC) and 
spin-on carbon (SOC) hard mask with an \ac{EBDW} system (SB3054, VISTEC) and subsequent plasma dry 
etching (Centura\,\textsuperscript{\textregistered}, Applied Materials) in the \SI{200}{\milli\meter} 
production line at CEA-Leti. The half-height diameters of nanopillars range from \SI{25}{\nano\meter} to 
\SI{50}{\nano\meter} and the pitch is larger than \SI{250}{\nano\meter} thus redeposition of sputtered 
atoms is prevented. The size of a nanopillar array is \SI{5}{\micro\meter}$\times$\SI{5}{\micro\meter}. 

The nanopillars are irradiated with a \SI{25}{\kilo\electronvolt} focused Ne$^+$ ion beam from the 
\ac{HIM}\@. Scanning the beam over a small area of \SI{5}{\micro\meter}$\times$\SI{2}{\micro\meter} 
with a sufficiently small pixel spacing and dwell 
time emulates the conditions in a broad-beam implanter. The beam current is restricted under 
\SI{500}{\femto\ampere} using a \SI{20}{\micro\meter} Au aperture. Under this condition the 
time interval between two arriving Ne ions is longer than \SI{320}{\nano\second} compared to the typical 
lifetime of the collision cascade which is on the order of tens of \SI{}{\pico\second}, thus 
preventing temporal overlap of the collision cascades. 
To perform irradiations at elevated temperatures in the \ac{HIM} a home-built heater stage with a 
tungsten filament capable of reaching \SI{\sim 500}{\celsius} is employed. 
The temperatures of 
the sample and the stage are monitored with type K thermocouples. Unless noted otherwise the most 
important remaining \ac{HIM} Ne$^+$ irradiation conditions are: \SI{1}{\nano\meter} pixel 
spacing and \SI{0.1}{\micro\second} dwell time. 
Images of the nanopillars are taken using a \SI{25}{\kilo\electronvolt} He$^+$ 
beam at 85\textdegree{} tilt angle to sample surface normal with a probe size smaller than 
\SI{0.5}{\nano\meter}. In addition, broad beam irradiation has been carried out using 
\SI{50}{\kilo\electronvolt} Si$^+$ from a \SI{200}{\kilo\electronvolt} ion implanter
(Danfysik Model 1090) at the Ion Beam Center (IBC) of the
Helmholtz-Zentrum Dresden-Rossendorf (HZDR)

Cross-sectional samples to analyse the lateral shape of selected Si nanopillars were obtained by FIB 
(Zeiss NVision 40) milling and lift-out of \SI{1}{\micro\meter}
$\times$\SI{50}{\micro\meter} TEM lamellae (about \SI{50}{\nano\meter} in thickness). 
Images have been acquired with an FEI Titan 80--300 microscope using \ac{bftem} with a 
contrast-enhancing objective aperture and \ac{eftem}
using the plasmon-loss peak at \SI{17}{\electronvolt} to show the contrast of Si abundance. 
These data allow to analyze the crystallinity of the nanopillars and the dimensions can be 
measured even when the pillar is amorphized. Nanopillar diameter and height,
as well as the volume reconstruction for the sputter yield calculation were achieved with the help 
of software Fiji~\cite{Schindelin2012}.

\ac{MC} codes based on the \ac{BCA} have been employed to simulate 
the ballistic processes during ion irradiation in order to contribute to the interpretation of 
the experimentally observed phenomena. The static program TRI3DST~\cite{Nietiadi2014,Urbassek2015} 
models the irradiation of three-dimensional (3D) bodies whose surface can be described by analytical 
functionals. TRI3DYN~\cite{Moeller2014,Moeller2016} is employed for fully 3D dynamic simulations 
of the modification of an irradiated structure during bombardment, with arbitrary bodies being 
set up in a 3D voxel grid. Both codes assume amorphous materials and do not include collective 
phenomena such as the viscous flow. Details of the simulations are given in the Supplementary Information.

\section{Results and Discussions}

\subsection{Amorphization of Si nanopillars}

\begin{figure}[tb]
   \centering
   \includegraphics[width=0.5\textwidth]{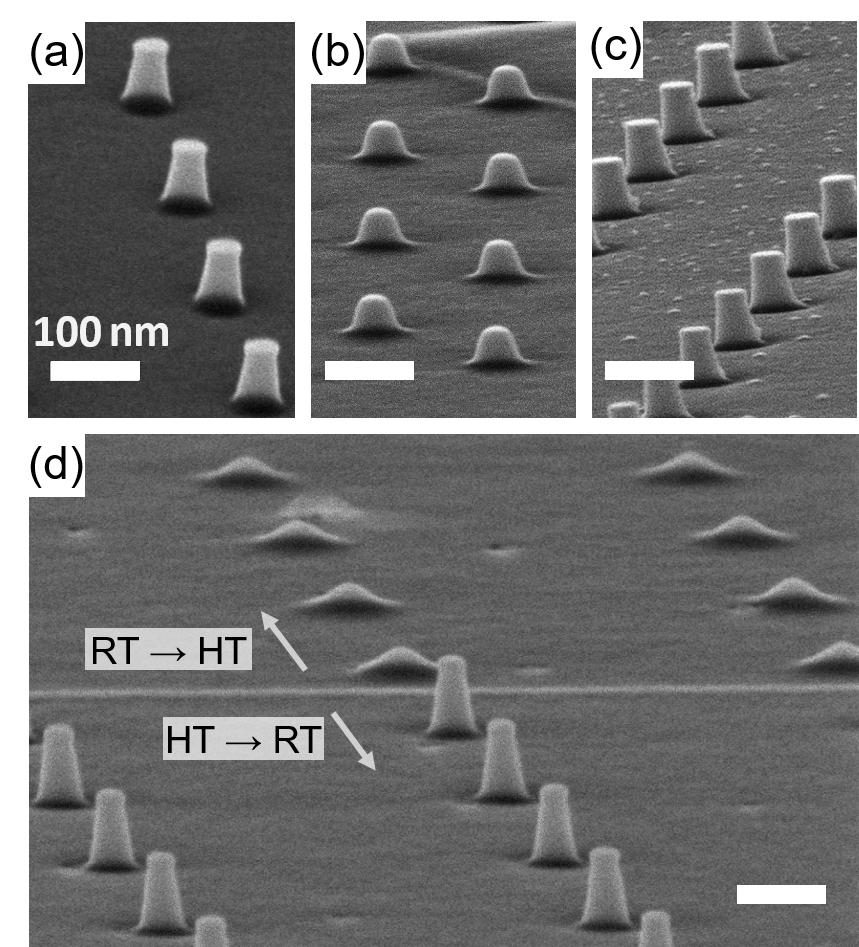}
   \caption{\label{fig:1} 
   	  Si nanopillars with an original diameter of \SI{25}{\nano\meter} have been irradiated 
   	  with \SI{25}{\kilo\electronvolt} Ne$^+$ and subsequently imaged at 85\textdegree{} with 
   	  He$^+$ from a \ac{HIM}. Image (a) shows the nanopillars before irradiation 
      while (b) and (c) show the nanopillars after \SI{2d16}{\per\centi\meter\squared} Ne$^+$ irradiation at \ac{RT}
      and \SI{400}{\celsius}, respectively. Image (d) shows the direct comparison of nanopillars after 
      having received \SI{2d15}{\per\centi\meter\squared} at room-temperature and 
      \SI{1.8d16}{\per\centi\meter\squared} at \SI{400}{\celsius} but in different order. Scale bars 
      in all images indicate \SI{100}{\nano\meter}.}
\end{figure}

Selected silicon nanopillars are irradiated from the top at both \ac{RT} and \ac{HT} (\SI{400}{\celsius}) 
with \SI{25}{\kilo\electronvolt} \neion{} using a fluence of \SI{2d16}{\per\centi\meter\squared}. 
\ac{HIM} images recorded using an \SI{85}{\degree}-tilt of the non-irradiated, \ac{RT}-irradiated and 
\ac{HT}-irradiated nanopillars are shown in Figure~\ref{fig:1}(a)-(c), respectively. 
After the irradiation at \ac{RT}, a strong change in shape of the nanopillars---characterized by a 
rounding of the upper edge accompanied by a severe loss of the height---can be observed. 
However, no such shape change is seen in the case of \ac{HT} irradiation of the nanopillars. 

In Fig.~\ref{fig:1}(d) identical nanopillars are irradiated with a combined \ac{RT} and \ac{HT} irradiation but in a different sequence. 
The resulting morphology can be compared in a qualitative way in the image. 
Both pillar fields have received the same total fluence but in reverse orders.
The pillar field in the back has first been irradiated at \ac{RT} with a fluence of only 
\SI{2d15}{\per\centi\meter\squared} followed by a fluence of \SI{18d15}{\per\centi\meter\squared} 
at an elevated temperature of \SI{400}{\celsius}. 
The fluence of \SI{2d15}{\per\centi\meter\squared} for the first irradiation step has been chosen 
so that it leads to an amorphization of the pillars. 
To ensure a complete amorphization of the pillars the chosen fluence is about 5 times higher than what has been reported for the amorphization of silicon by few \SI{10}{\kilo\electronvolt} Ne$^+$ irradiation~\cite{Dennis1978}.
However, the fluence is small enough to not lead to observable changes in the morphology of the pillars.
The pillars in the foreground of Fig.~\ref{fig:1}(d) were first irradiated at \SI{400}{\celsius} 
with a fluence of \SI{18d15}{\per\centi\meter\squared} followed by \SI{2d15}{\per\centi\meter\squared} 
after being cooled down to \ac{RT}.
As a result both pillar fields have received a total fluence of \SI{2d16}{\per\centi\meter\squared}.
However, as is clear from Fig.~\ref{fig:1}(d) the sequence of \ac{RT} vs. \ac{HT} irradiation plays 
an important role for the final morphology. 
The amorphous pillars in the background---irradiated first at \ac{RT}---are nearly completely removed. 
A closer look also reveals that the diameter at the foot of the former pillars has increased.
The pillars in the foreground---irradiated first at \ac{HT}---still show their pristine shape with 
nearly no observable change in morphology.
The final irradiation of the foreground pillars at \ac{RT} ensures they receive the same amount of amorphization 
as the background pillars received initially. Again, this does not lead to a noticeable change of the pillar shape.

\begin{figure}[tb]
   \centering
   \includegraphics[width=0.5\textwidth]{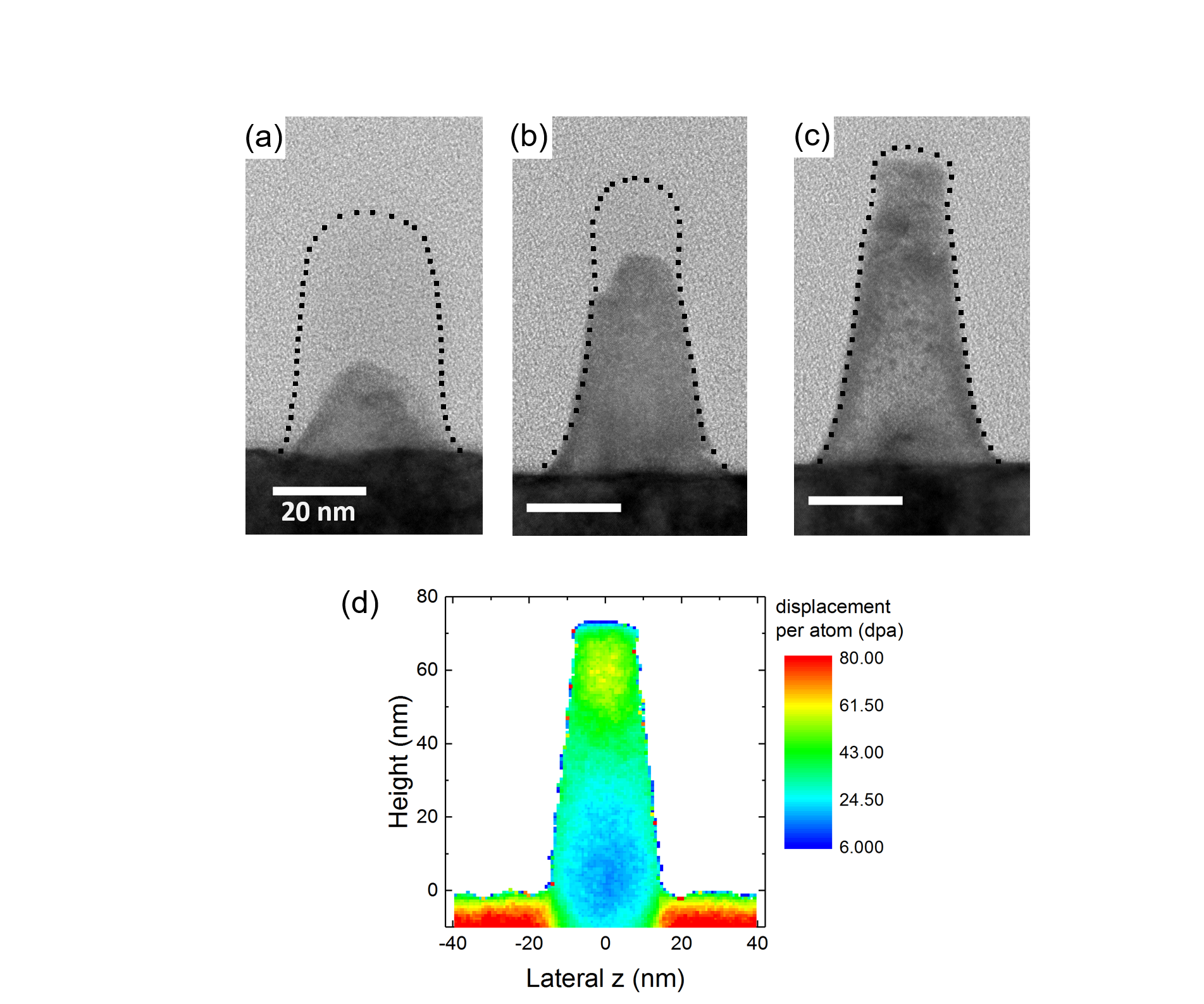}
   \caption{\label{fig:2}\ac{bftem} micrographs of Si nanopillars irradiated with 
		   	\SI{2d16}{\per\centi\meter\squared}, \SI{25}{\kilo\electronvolt} \neion{} at (a) \SI{250}{\celsius}, 
		   	(b) \SI{300}{\celsius} and (c) \SI{350}{\celsius} showing the cases of the nanopillar being almost 
		   	entirely amorphized, only top segment amorphized and entirely crystalline. The black dashed lines 
		   	indicate the outlines of the actual Si nanopillars. (d) TRI3DYN simulation 
		   	showing the distribution of displacements created by \SI{25}{\kilo\electronvolt} \neion{} of 
		   	\SI{2d16}{\per\centi\meter\squared} fluence. The profile has to be compared to the extent of the 
		   	crystalline and amorphous regions after an irradiation at temperatures lower than T$_\mathrm{c}$.
   }
\end{figure}

A detailed investigation of the temperature dependence of the ion beam induced amorphization is carried out 
on similar structures. A series of nanopillars is irradiated with Ne$^{+}$ ions using a fluence of 
\SI{2d16}{\per\centi\meter\squared} at various temperatures from \SI{250}{\celsius} up to \SI{350}{\celsius} 
in \SI{25}{\celsius} steps. 
In Fig.~\ref{fig:2}(a)-(c) selected \ac{bftem} micrographs of typical structures are presented (\ac{TEM} 
images for the other temperatures can be found in the supporting information). 
In BF-TEM micrographs only the crystalline part of the nanopillar is visible, and it is evident that after low 
temperature irradiation the pillar becomes amorphized. The black dashed lines in Fig.~\ref{fig:2}(a)-(c) 
outline the border of the actual pillar as extracted from corresponding \ac{eftem} micrographs (not shown 
here; please see the supporting information for examples). 
With increasing irradiation temperature, at the same fluence an increasing part of the pillar stays crystalline and only a 
small part of the pillar becomes amorphous. Finally, for an irradiation temperature of \SI{350}{\celsius} 
no amorphization can be detected from the \ac{bftem} images. From the investigation of the amorphization 
behaviour at additional temperatures (see the supporting information), we conclude that the critical 
temperature of amorphization (T$_\mathrm{c}$) during \SI{25}{\kilo\electronvolt} \neion{} ion irradiation 
for the presented Si nanopillars lies between \SI{325}{\celsius} and \SI{350}{\celsius}. 

With the \ac{MC} simulation program TRI3DYN, the dynamic evolution of a Si nanopillar under ion irradiation 
can be visualized. In Fig.~\ref{fig:3}(d) the accumulated \ac{dpa} of the central slice of a nanopillar 
after \SI{25}{\kilo\electronvolt} Ne$^{+}$ irradiation with a fluence of \SI{2d16}{\per\centi\meter\squared} 
is shown. In this particular case, the comparison between the simulation and experiments is a simplification 
from the complex interplay leading to amorphization, including the original \ac{dpa} created by the 
ion trajectory, temperature-dependent dynamic annealing, overlapping defect pockets as well as 
deformation due to viscous flow. \ac{BCA} type \ac{MC} simulation programs like TRI3DYN usually 
only treat amorphous samples, and effects such as dynamic annealing and viscous flow are not considered. 

The obtained simulation results can well explain the amorphization profiles obtained by \ac{bftem} presented 
in Figs.~\ref{fig:2}(a,b). In these cases, only the top of the pillar is amorphized which corresponds to 
the region with the highest defect density according to the simulation result presented in Fig.~\ref{fig:2}(d). 
Furthermore, from the \ac{bftem} micrographs in Figs.~\ref{fig:2}(a,b) one can see that the interface between 
the amorphous and crystalline part of the Si nanopillar tends to bend towards the bottom of the pillar. 
This also agrees well with the simulated result which shows that in the lower part of the nanopillar the 
displacement density is higher close to the sidewalls than in the centre. 
This is attributed to the slightly tapered sidewall of the nanopillars (typically \SI{7}{\degree}) which 
leads to high angle sputtering, creating high amount of displacements close to the pillar surface. 
From the TRI3DYN result presented in Fig.~\ref{fig:2} we conclude that a fluence of 
\SI{2d16}{\per\centi\meter\squared} at \ac{RT} results in a damage of at least \SI{6}{\ac{dpa}}. 
This is more than sufficient to amorphize the entire nanopillar in the absence of any dynamic annealing processes. 

According to the defect out-diffusing model of amorphization~\cite{Morehead1970}, when the tem\-per\-a\-ture 
of the crystalline silicon is higher than the amorphization critical temperature T$_\mathrm{c}$, no matter 
how high the fluence is during an irradiation the originally crystalline structure will not be amorphized. 
From Figs.~\ref{fig:1}(c,d) and~\ref{fig:2}(c) one can see that during irradiation at temperatures higher 
than T$_\mathrm{c}$ the pillars stay crystalline and no morphological changes can be observed.
This temperature of \SI{325}{\celsius} to \SI{350}{\celsius} is still \SI{200}{\celsius} lower than the 
regime where epitaxial recrystallization could recover a large-area amorphous structure~\cite{Crowder1970}.
The nanopillars irradiated with \SI{25}{\kilo\electronvolt} \neion{} at \ac{RT} become amorphous already 
after a very low fluence, similar to the amorphization fluence in bulk Si which is between 
\SI{1d15}{\per\centi\meter\squared} and \SI{2d15}{\per\centi\meter\squared}~\cite{Dennis1978}.
While the dynamic annealing by the ion beam at \ac{HT} above T$_\mathrm{c}$ is sufficient to prevent 
amorphization of the initially crystalline nanopillars, it can not recover the crystal structure of an initially 
amorphous nanopillar by epitaxial recrystallization. 
As a consequence, surface tension induced by ion irradiation leads to viscous behavior of amorphized 
nanostructures and the resulting plastic deformation visible in Fig.~\ref{fig:1}(b) and in the back 
of Fig.~\ref{fig:1}(d), as shown in previous studies---albeit mainly at ion energies from \SI{100}{\kilo\electronvolt} 
to a few \SI{}{\mega\electronvolt}---with various ion species, energies and target materials~\cite{volkert1991,van2003,Johannes2015a}.


To summarize, we showed that above the critical temperature of amorphization T$_\mathrm{c}$ dynamic 
annealing prevents the amorphization of the crystalline Si nanopillars. 
From the \ac{TEM} analysis we find out that this temperature corresponds to \SI{325}{\celsius} to 
\SI{350}{\celsius} for irradiation with \SI{25}{\kilo\electronvolt} \neion{}.
This is in agreement with the results from \ac{HIM} imaging where we find severe deformation of 
initially crystalline nanopillars only after irradiation at temperatures below T$_\mathrm{c}$.
This deformation is a result of viscous flow of the amorphous material due to capillary forces 
acting on the nanoscale pillars. Initially amorphous pillars experience dramatic shape changes 
even if irradiated at \ac{HT}, as the temperature for epitaxial recrystallization is still 
\SI{200}{\celsius} higher and dynamic annealing is not sufficient to recrystallize the nanopillars.

\subsection{Thinning of the nanopillars}

\begin{figure}[tb]
	\centering
	\includegraphics[width=0.5\textwidth]{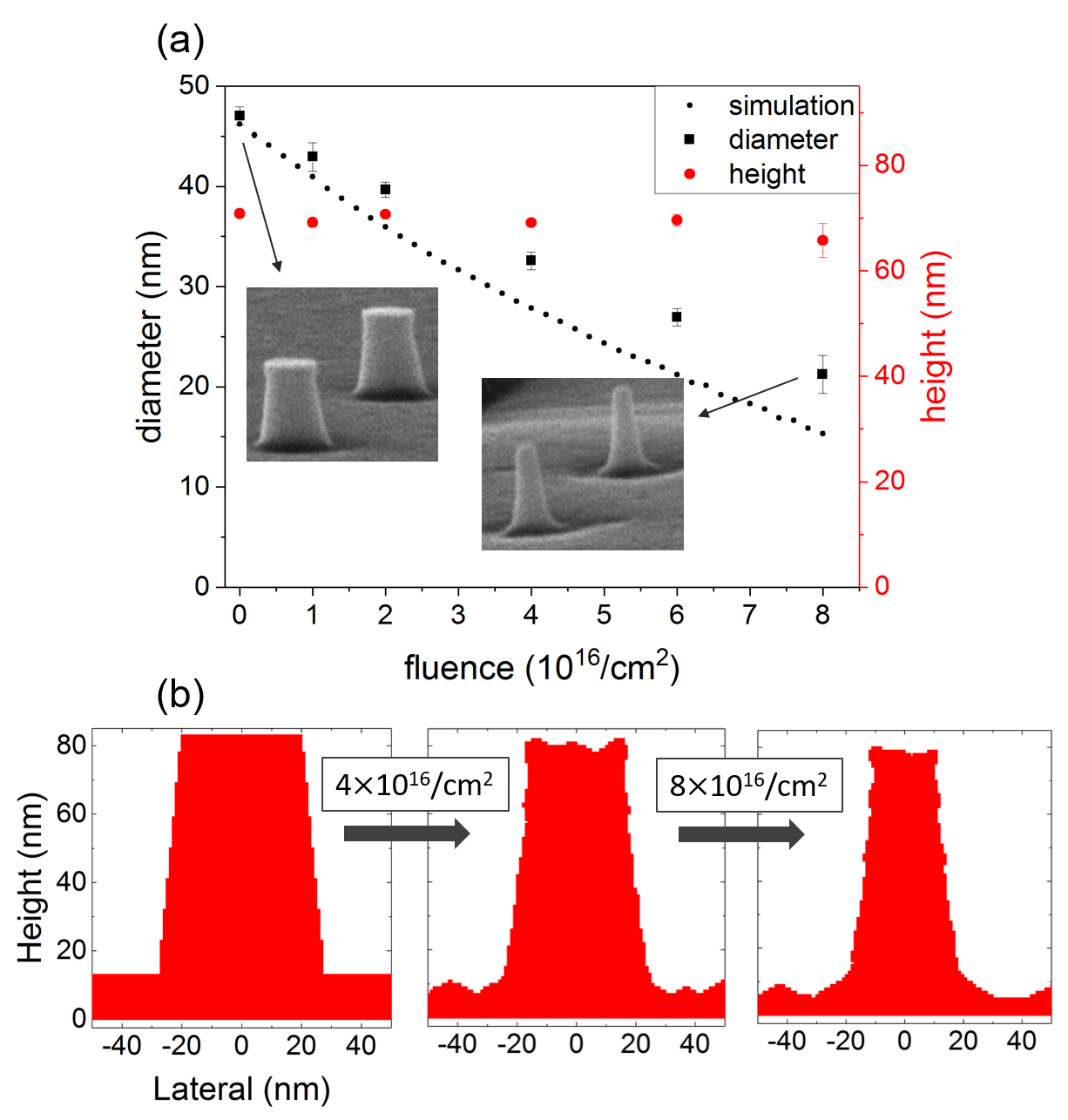}
	\caption{\label{fig:3} (a)Si nanopillar irradiated at \SI{400}{\celsius} with 
		fluence up to \SI{8d16}{\per\centi\meter\squared} shows a steady decrease of diameter 
		without noticeable shrinkage of height. The nanopillars before and after 
		irradiation are imaged with \ac{HIM} at 85\textdegree{} tilt angle and shown 
		as insets. The thinning process is also simulated with TRI3DYN and the diameter is plotted 
		as an comparison to the experimental results. (b)Evolution of the central 
		\SI{3}{\nano\meter} slice of the simulated nanopillars during the Ne$^+$ 
		thinning process. The diameters at half-height are plotted in (a) showing 
		good agreement compared to the experimental results.}
	
\end{figure}

Si nanopillars were irradiated at a temperature well above T$_\mathrm{c}$ to fully exclude the influence of ion-induced amorphization, and the applied fluence has been increased up to \SI{8d16}{\per\centi\meter\squared}. The nanopillars studied here have 
a diameter at half-height of \SI{47}{\nano\meter} and height of \SI{71}{\nano\meter}. 
After the irradiation, the sample was immediately imaged 
with the \ac{HIM} at a tilt angle of 85\textdegree{}. In Fig~\ref{fig:3}(a) the height and 
diameter of the nanopillars with the same original diameter after different irradiation fluences 
are plotted. With the increase of the fluence, a steady 
decrease of the diameter has been observed while the height remains almost unchanged. Linear fitting of the diameter reduction 
shows a slope, i.e.\ the reduction rate of \num{-3.3(1)}\,\SI{}{\nano\meter}/($1\times10^{16}$\SI{}{\per\centi\meter\squared}). 
In the same diagram the 
simulated results of the thinning process from the program TRI3DYN are shown. Snapshots of the central \SI{3}{\nano\meter} slice of the nanopillar after different irradiation fluences are presented in Fig.~\ref{fig:3}(b). 
While the simulation qualitatively fits with 
the experimental data it slightly overestimates the sputter yield for smaller diameters. Such a tendency was also observed in previous work~\cite{Johannes2015a} where the ion irradiation 
was performed at 45\textdegree{} incidence relative to the axis of a Si nanowire. 
The smaller the pillar diameter is, 
or when the ion impact position is close to the nanopillar rim, the higher the chance 
will be that a recoiled atom will leave the structure with a high kinetic energy. 
This reduces the energy deposited inside the pillar and subsequently leads to less energy deposited per incident 
ion as compared to a larger diameter pillar or a bulk system.

\begin{figure}[tb]
	\centering
	\includegraphics[width=0.5\textwidth]{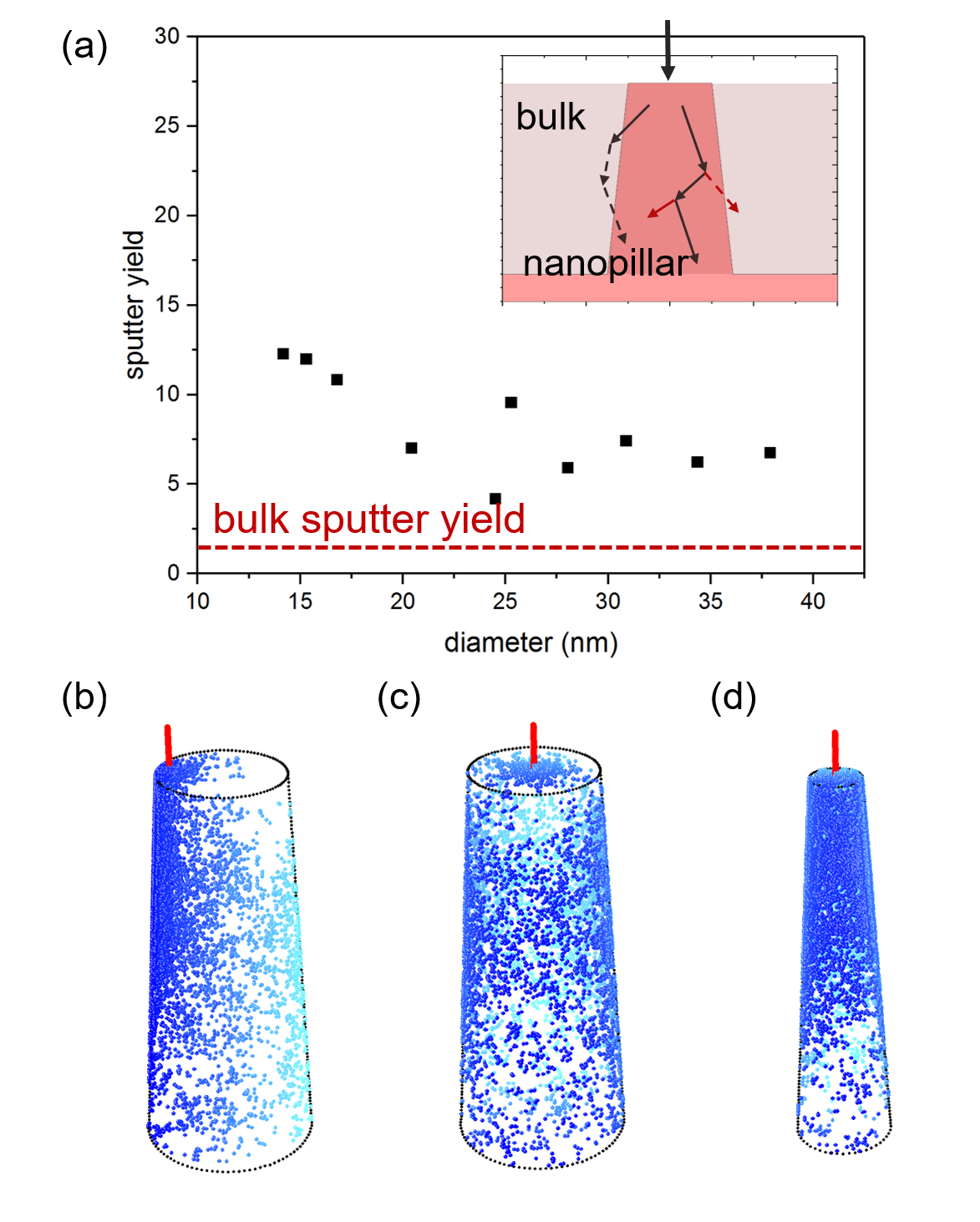}
	\caption{\label{fig:4}(a) Experimentally obtained sputter yield for \SI{25}{\kilo\electronvolt} 
		Ne$^+$ in Si nanopillars plotted against the diameters at half height of the nanopillars. Due to enhanced forward sputtering 
		the sputter yield in nanopillars is at least 3 times higher than the sputter yield in the bulk. 		
		(b)-(d) TRI3DST simulations showing profiles of surface sputtering events enhanced forward 
		when an incident ion is (b) 
		close to the pillar rim, (c)in the centre of the pillar or (d)when the pillar diameter is comparable 
		with the size of the collision cascade. The diameters of the nanopillars in (b)-(d) 
		are \SI{30}{\nano\meter}, \SI{30}{\nano\meter} and \SI{15}{\nano\meter}, respectively. The colour 
		coding (from dark blue to light blue) indicates the position of a sputtering event in the horizontal direction.}
\end{figure}

To further analyze the mechanism of the decrease of the nanopillar diameter, the average 
sputter yield during the pillar thinning process has been measured using the incident ion current 
and the lateral dimensions of the pillars and plotted against the diameter 
of the pillars. Detailed description of calculating the sputter yield from \ac{TEM} 
micrographs is included in the Supporting Materials.  
As shown in Fig.~\ref{fig:4}(a), the average sputter yield is approximately constant as long as 
the diameter of the 
nanopillars is significantly larger than the lateral range of the Ne$^+$ ions of \SI{\sim{}19}{\nano\meter}.
The sputter yield measured for the nanopillars is 
at least 3 times higher than what is expected for sputtering at normal incidence on bulk Si surfaces as 
obtained from TRI3DST or SRIM\@. 

Two factors are contributing to such a high sputter yield. First, due to the approximately 7\textdegree{}
tapering, the sidewalls of the nanopillar are also exposed to the irradiation at a high glancing angle 
which results in a strongly enhanced sputter yield. This contribution is hard to quantify
due to the additional radial curvature of the nanopillar sidewall. The second contribution comes from 
is the forward sputtering from ions hitting at normal incidence on the top of the nanopillar. 
From the schematic presented as an inset in Figure~\ref{fig:4}(a) one can see that in contrast to the 
case of a bulk structure (indicated as light red) where the collision cascade is fully embedded inside 
the sample, in the case of a nanopillar, the lateral extent of the collision cascade has a high 
probability to overlap with the surface. When a recoiled atom has a kinetic energy higher 
than the surface binding energy for Si of \SI{4.7}{\electronvolt}~\cite{Nietiadi2014},
it will leave the nanopillar as a forward sputtered atom. Detailed analysis from the simulation results in 
Fig.~\ref{fig:3}(b) shows that at the initial diameter approximately \SI{74}{\percent} of the 
sputtering can be attributed to the forward sputtering from 
the ions penetrating into the top of the nanopillar. 

As the irradiation proceeds, the two contributions evolve in 
different manners. The high-angle sputtering on the sidewalls of the nanopillars 
slightly decreases with the decreasing diameter and unchanged sidewall tapering (see Fig.~\ref{fig:3}(b) and 
the insets in Fig.~\ref{fig:3}(a)). On the other hand, forward sputtering from ions hitting the 
top of the pillar will be more severe due to a larger surface to volume ratio. 

Results from simulations, using TRI3DST, of focused-beam 
irradiation onto a Si nanopillar are shown in Fig.~\ref{fig:4}(b)-(d). 
The ions hit the top surface of the nanopillar at (b) \SI{5}{\nano\meter} from the rim and (c)-(d) 
in the centre. For statistical reasons, an incidence of 1000 ions was simulated in all cases. In Fig.~\ref{fig:4}(b)-(d), 
blue dots indicate the position on the nanopillar sidewall where a sputter event occurred. It 
is clear that for every single incident ion, the smaller 
the diameter of the nanopillar is, or the closer the incident spot is located to the rim, the 
higher the amount of sputtered atoms will be. This behavior results in a steady 
decrease of the nanopillar diameter, mostly due to the enhanced forward sputtering. In this 
model, two distinctive stages will be observed for such a thinning process. When the diameter 
of the nanopillars is significantly larger than the lateral 
range of the collision cascade, the incident ions that create forward sputtering near the rim, such 
as in the case of Fig.~\ref{fig:4}(b), are unlikely to sputter from the other side of the nanopillar. 
Incident ions closer to the centre of the nanopillar, as shown in Fig.~\ref{fig:4}(c), would only 
induce a small amount of forward sputtering as the collision cascade barely reaches the pillar surface. 
As a consequence, the decrease of the pillar diameter has a linear dependency on the fluence. 
When the diameter of the nanopillars is comparable to or smaller than the size of the 
collision cascade, which is \SI{18.6}{\nano\meter} in the case of \SI{25}{\kilo\electronvolt} Ne$^{+}$, 
the chance that a recoil atom leaves 
the nanopillar via forward sputtering is high, independent from the ion impact position. 

However, with decreasing pillar diameter, the size and shape of the collision cascade will also change. 
This is due to the limited size of the nanopillar, which reduces the actual average lateral range of the ions to a 
value smaller than the diameter but also reduces the range of the ions as they are more likely to leave 
the target structure. This situation is depicted in the inset of Fig.~\ref{fig:4}(a). The lower part of the bulk 
collisions cascade is missing in the collision cascade inside the nanopillar.
This also result in a redistribution of the energy deposition and the related defect density. In the nanopillar 
more energy is deposited close to the top which leads to more sputtering at the top and eventually a reduction of 
the pillar height.

\begin{figure}[tb]
   \centering
   \includegraphics[width=0.5\textwidth]{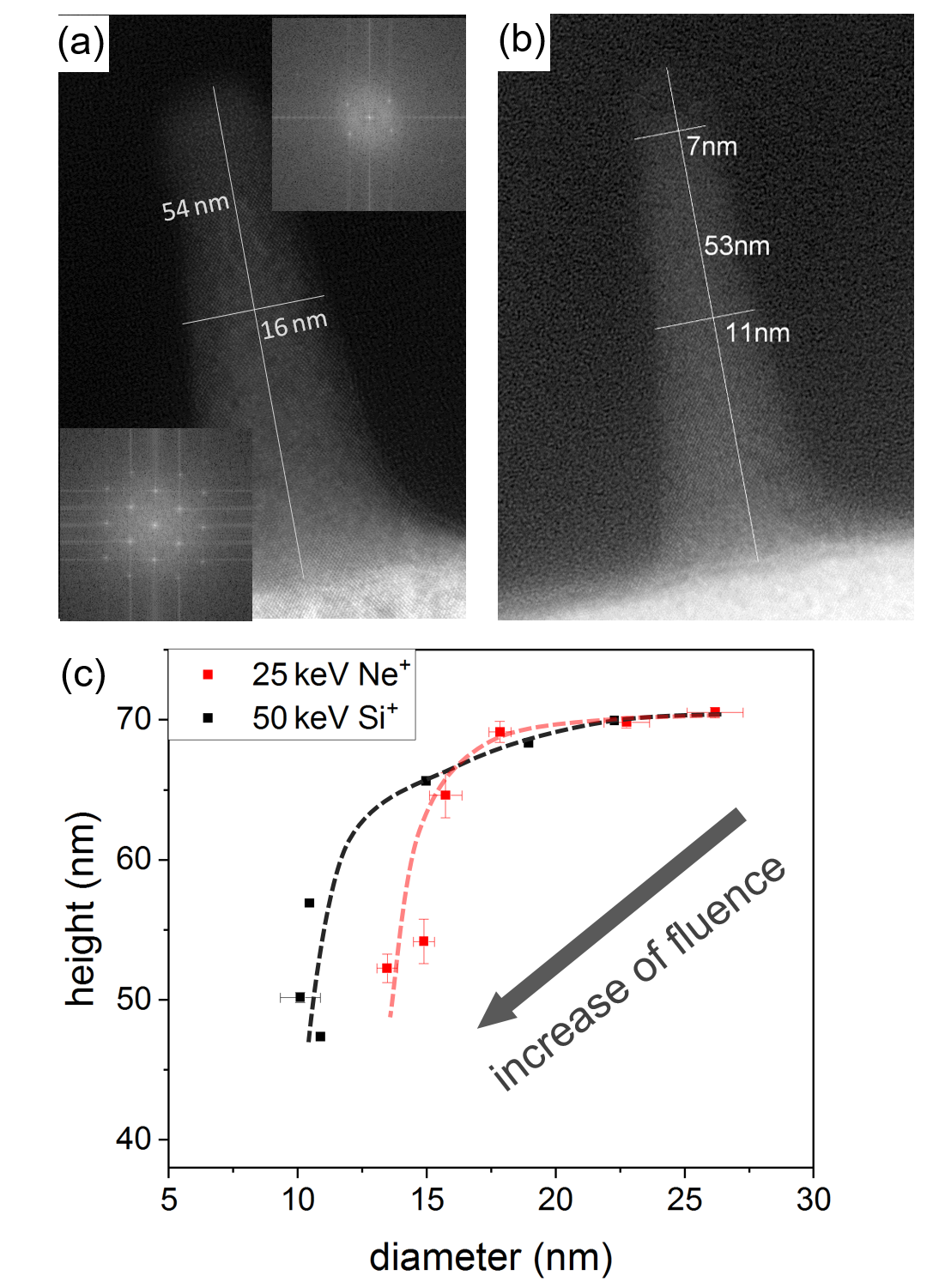}
   \caption{\label{fig:5}\ac{eftem} micrographs show the nanopillars after HT-irradiation 
   	  with (a)Ne$^+$ and (b)Si$^+$ of \SI{6d16}{\per\centi\meter\squared} fluence. Fast Fourier 
   	  transform (FFT) shows in both cases the top segment of the nanopillar is partially 
   	  amorphized while the lower part remains crystalline. (c) As measured from 85\textdegree{}-tilted 
      \ac{HIM} and \ac{eftem} images, the thinning from Ne$^+$ in Si nanopillars is limited when the 
      diameter gets smaller than 16\,nm while using Si$^+$ for the same purpose could reduce the 
      diameter to \SI{11}{\nano\meter}. }
\end{figure}

To demonstrate the potential for upscaling to industrial applications, we also performed Si$^+$ broad 
beam irradiation of nanopillar arrays in a standard ion implanter. To be able to compare the results to the previous 
\neion{} irradiation, we used \SI{50}{\kilo\electronvolt} Si$^+$, which has the similar dpa as \SI{25}{\kilo\electronvolt} Ne$^+$, to irradiate the nanopillars 
at \SI{400}{\celsius}. 
In Fig.~\ref{fig:5}(a) and (b) \ac{eftem} micrographs of nanopillars 
after a fluence of \SI{6d16}{\per\centi\meter\squared} with focused Ne$^+$ and broad-beam 
Si$^+$ are shown. 
These experimental conditions were chosen as they represent the highest fluences applied in 
both irradiation conditions that do not lead to a strong decrease of the nanopillar height. 
Both cases indicate the experimental limit for possible lateral pillar shrinking. \ac{fft} results 
of the upper and lower part of an irradiated nanopillar 
are shown as insets in Fig.~\ref{fig:5}(a). They reveal that while the lower part of the Si 
nanopillar remains single crystalline the top part is characterized by a 
co-existence of crystalline and amorphous areas. As we further increase the irradiation fluence, 
the height of the nanopillar start to decrease and the interface between the crystalline and 
amorphous parts moves towards the bottom. In the case of Si nanopillars irradiated with 
Si$^+$ of \SI{6d16}{\per\centi\meter\squared} fluence, the majority of the nanopillar also remains 
crystalline despite the higher density of defects in the vicinity of the top surface, indicating 
a better structural integrity than the one irradiated with Ne$^+$.

In Figure~\ref{fig:5}(c) the average height and diameter of the nanopillars during the thinning 
processes via focused Ne$^+$ and 
broad-beam Si$^+$ are plotted. The dashed lines are guides to the eyes to allow a clear comparison 
of the trends between 
irradiation with the two ion species.
In the case of Si$^+$ irradiated 
nanopillars, the diameter was measured via 85\textdegree-tilted \ac{HIM} imaging subtracting the thickness 
of native oxide on the nanopillar sidewalls. While in both cases the curves eventually turn steep 
as the height 
starts to decrease and the thinning process slows down, allowing for a \SI{15}{\percent} reduction of 
height, the achievable diameters 
for Ne$^+$ and Si$^+$ thinned nanopillars are \SI{14}{\nano\meter} and \SI{11}{\nano\meter}, respectively. 

The difference in final diameter may be attributed to the following factors. 
First, with the projected ranges R$_\mathrm{p}$ of 73.2\,nm for 50\,keV Si$^+$ and 56.9\,nm 
for 25\,keV Ne$^+$, the Si$^+$ 
better fits the height of the original nanopillars. 
A higher energy of the incident ion would result in a deeper and shallower energy deposition thus postponing
the defect accumulation at the top segment of the nanopillars. Second, Si self-irradiation leads to a more 
forward directed collision cascade due the optimal energy transfer between the incident Si and the target Si 
atoms having the same mass. As a consequence Si irradiation results in more homogeneous sputtering of the 
pillar and a slower reduction of the longitudinal range due cascade truncation by the pillar.

\section{Conclusion}

In this work, nanopillars with diameters of \SI{50}{\nano\meter} and below were irradiated with 
Ne$^+$ from a helium ion microscope at temperatures between \SI{250}{\celsius} and 
\SI{400}{\celsius}. The critical temperature of amorphization under ion beam irradiation 
T$_\mathrm{c}$ was found in the range between \SI{325}{\celsius} and \SI{350}{\celsius}. The morphology of the Si 
nanopillars evolves accordingly at different temperatures during Ne$^+$ irradiation. At temperatures lower than T$_\mathrm{c}$, the 
nanopillar shape turned conical and a strong decrease in height occurred due to the presence of viscous flow. 
When irradiated at temperatures higher than T$_\mathrm{c}$, the Si nanopillars remained crystalline and the viscous flow did not occur, while the diameter of the nanopillars was reduced linearly with the applied fluences. Such a thinning 
process was simulated and visualized via the static simulation programme TRI3DST\@. The sputter yields 
extracted from both experiments and simulation indicate an enhanced sputtering 
effect on the nanopillars which can be attributed to the high surface-to-volume ratio and the overlap between 
the collision cascade and the nanopillar surfaces. In particular the latter leads to a truncation of the 
collision cascade and an increasing contribution from forward sputtering to the thinning process. 
However, the truncation of the collision cascade in particular in the longitudinal direction increases the 
fraction of energy deposited in the top part of the pillar, resulting in an increase of the sputter yield 
in this region and ultimately leads to a reduction in height. 

Compared to earlier attempts of ion beam based pillar reduction~\cite{Johannes2015a} our method does not require 
rotation of the incident direction and results in smaller final diameter.
Our approach which utilizes an irradiation process at slightly elevated but still \ac{VLSI} compatible temperatures 
enables new possibilities for the fabricating of vertical nanostructures, such as pillars and fins, in a 
top-down and \ac{cmos}-compatible manner which is not limited by the lithographic resolution.

\ack{}
This work has been funded by the European Commission H-2020 programme ``IONS4SET''
under grant agreement No.\,688072. The authors thank Roman B\"ottger, 
Ulrich Kentsch and the Ion Beam Center (IBC) at HZDR for the broad-beam irradiation.

\bibliography{pillar-thinning}
\bibliographystyle{unsrt}




\section*{Supplementary material}

# Supporting Information— Morphology modification of Si nanopillars under ion irradiation at elevated temperatures: plastic deformation and controlled thinning to 10 nm

Xiaomo Xu, Karl-Heinz Heinig, Wolfhard Möller, Hans-Jürgen Engelmann, Nico Klingner, Johannes von Borany and Gregor Hlawacek‡

Institute of Ion Beam Physics and Materials Research, Helmholtz-Zentrum Dresden-Rossendorf, Bautzner Landstrasse 400, 01328 Dresden, Germany

Ahmed Gharbi, Raluca Tiron

CEA-Leti, Rue des Martyrs 17, 38054 Grenoble, France

E-mail: g.hlawacek@hzdr.de

## 1   Ne$^+$ amorphization of nanopillars

Figure S1 shows energy-filtered transmission electron microscopy (EFTEM) and bright-field transmission electron microscopy (BF-TEM) micrographs of nanopillars irradiated at (a-b) 275 °C and (c-d) 325 °C with 25 keV Ne$^+$ of $2 \times 10^{16}$ cm$^{-2}$ fluence. By comparing the EFTEM and BF-TEM images it is clear that the majority of a nanopillar after the irradiation at 275 °C is amorphous while nearly the entire nanopillar after the irradiation at 325 °C remains crystalline. Together with Figure 2 we conclude that the critical temperature of amorphization in this case is slightly higher than 325 °C.

## 2   Experimental measurement of the sputter yield

The sputter yield on a nanopillar can be calculated via a series of EFTEM micrographs of originally identical nanopillars after different fluencies. According to the definition, the sputter yield is calculated by dividing the total number of sputtered atoms (N$_{\text{sputter}}$) with the total number of ion hitting the nanopillar (N$_{\text{ion}}$). Figure S2 shows an example

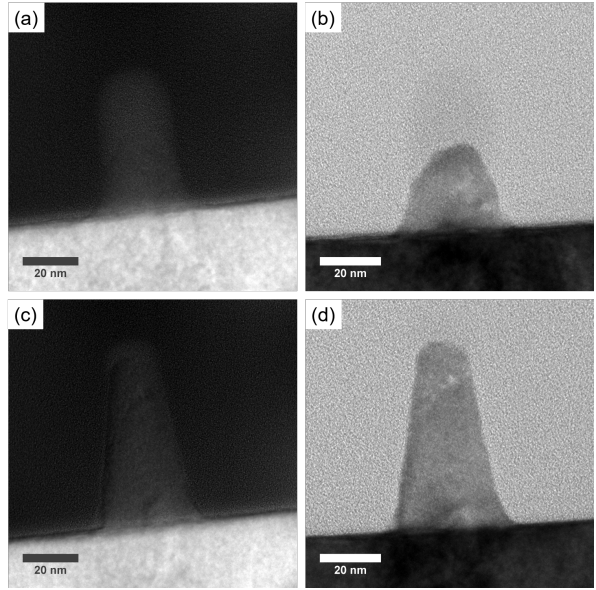

Figure S1: EFTEM and BF-TEM micrographs show nanopillars irradiated at (a-b) 275 °C and (c-d) 325 °C with 25 keV Ne$^+$ of $2 \times 10^{16}$ cm$^{-2}$ fluence. The majority of a nanopillar structure irradiated at 275 °C is amorphous while almost the entire nanopillar after the irradiation at 325 °C remains crystalline.

to measure N$_{sputter}$ and N$_{ion}$ with an original nanopillar, the same procedure was applied to nanopillars irradiated with varied fluences.

The nanopillar is imaged with EFTEM at 17 eV plasmon-loss peak (a) and the micrograph is rotated and cropped according to the shape of the nanopillar (b). Although the nanopillar has a irregularly tapered sidewall, at any certain height the horizontal cross-section is taken as circular. As shown in Figure S2(c), a Minimum Error threshold [1] is applied to distinguish the Si elemental contrast of the nanopillar from the background. In Figure S2(d) a mesh with 5 nm$^2$ grid is overlayed and the diameter at each height can be measured. Thus, the volume of a nanopillar can be approximately calculated by summing up the volume of all conical segments, and the sputtered atom number N$_{sputter}$ in each fluence interval can be calculated by multiplying the volume difference and the atomic density of Si (50 atoms/nm$^2$).

The number of incident ions N$_{ion}$ can be obtained by multiplying the fluence interval with the nanopillar cross-section. The diameter used for the cross-section calculation is taken as the half-height diameter to be consistent with the pillar thinning process in Figure 3 and Figure 5.

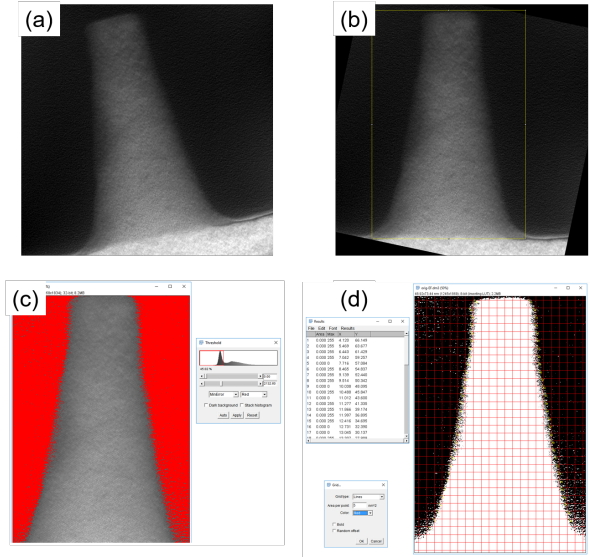

Figure S2: (a) An EFTEM micrograph shows a Si nanopillar before ion irradiation. After the image is rotated and cropped (b), a Minimum Error threshold is applied to distinguish the Si elemental contrast from the background. A mesh grid with the density of $5\,\text{nm}^2$ is overlayed on the nanopillar and the diameter and volume of a nanopillar can be measured.

## 3 Computer Simulations

For ballistic simulations of the scattering of impinging ions from the irradiated structure, their slowing down in the material and the effects of the associated collision cascades such as damage formation and surface sputtering, static TRI3DST [2, 3] and dynamic TRI3DYN [4, 5] codes have been employed which make use of the binary collision approximation (BCA). Their collisional algorithms, being valid for amorphous materials, are based on early static [6, 7] and dynamic [8, 9] versions of the widely spread TRIM code [10]. Briefly, the simulations trace the trajectories of incident ions and recoil atoms, which are generated in the associated collision cascades by binary atomic interactions in a screened Coulomb potential with the *Kr-C* parametrization [11]. Energy transfer to the target electrons is modeled by an equipartition of local [12] and nonlocal [13] energy losses. The results addressing surface sputtering and bulk atomic displacement depend on the preselected parameters of the surface binding energy and the displacement threshold energy, respectively. For the former, the standard selection of the sublimation enthalpy of $4.7\,\text{eV}$ for Si results in flat-surface sputtering yields which agree well with experimental data [14]. The definition of the displacement threshold energy is less straightforward. Tabulated data are available for crystalline materials at low damage

levels [15–17]. However, most materials are highly damaged under ion irradiation at high fluences and sufficiently low temperatures. Under the conditions of the present work, Si becomes amorphized already at fluences which are much smaller than the ones employed here. Then, displacements may occur at lower energies due to trapping at preexisting defects. In their pioneering theoretical paper on ion mixing, Sigmund and Gras-Marti [18] propose a threshold energy of 7.8 eV for cascade mixing in Si. In consistence with this choice, a default setting of $U_d$=8 eV has been successful in numerous ion mixing, preferential sputtering and thin film deposition studies using TRIDYN [19] and is also applied here.

In TRI3DST, a variety of three-dimensional bodies can be selected, which are defined by analytical functionals describing their surface. For the present paper (see Figure 4), conical Si nanopillars are irradiated in the direction along the pillar axis by a nanobeam with a Gaussian beam profile of 3 nm full-width at half maximum. For clarity of the extracted distributions of sputter ejection sites at simultaneously reasonable statistics, 1000 incident ions have been traced in each run.

TRI3DYN works with a fixed 3D grid of cuboidal voxels which span a cuboidal computational volume. On the voxel grid, arbitrary initial structures can be defined and their modification of the local bulk composition and surface contour during irradiation can be traced. For the present simulations (see Figures 2 and 3), tapered Si nanopillars of 70 nm height and upper/lower diameters of 19/29 nm (see Figure 2) and 42/56 nm (see Figure 3) have been set up on a planar Si substrate with voxel dimensions of 0.8×0.8×0.8 nm$^3$ and 1×1×1 nm$^3$, respectively. Periodic boundaries are applied in the directions normal to the pillar axes. Laterally uniform irradiation is accomplished with the ions starting at randomly selected positions on the upper plane of the computational volume. For the entry of ions at glancing incidence (in particular at the tapered flanks of nanopillars), the voxel structure may result in an artificially enhanced entry and trapping, whereas in reality such ions would be preferentially reflected from the wall [4]. Consequently, an algorithm of 3D surface planarization has been developed which replaces the cuboidal surface of each surface voxel by a locally planar surface with an inclination taking into account the neighboring surface voxels.

The ballistic transport of incident projectiles or recoil atoms generated in the collision cascades may remove atoms from specific voxels due to bulk displacements or surface sputtering, or add displaced atoms or slowed-down projectiles. (The incorporation of Ne projectiles is neglected in the present simulations as its contribution

23

to the local atomic densities is negligible in the present range of ion fluence.) For the dynamic relaxation of the structure during irradiation, each moving atom (*pseudoatom*) in the simulation represents a certain number of real atoms, which may be fractional (0.25 and 1.0 for the present cases of Figure 2 and Figure 3, respectively), and which is automatically chosen to minimize the computation time at still sufficient statistical quality of the dynamic development of the system. After a certain number of incident pseudoprojectiles each, the dynamic relaxation of the system is activated. In each modified bulk voxel, the atomic density is re-established. This is accomplished by material exchange between neighboring voxels and transport from/to surface voxels (for details, see Ref. [4]). Surface voxels which are depleted due to sputtering are combined and partially turned into vacuum voxels. Each relaxation procedure is terminated by an algorithm of surface smoothing.

## Acknowledgments



\end{document}